\documentclass[twoside,a4paper,11pt]{sea10}
\usepackage{graphicx}
\usepackage{hyperref}
\usepackage{movie15}
\begin{document}
\pagenumbering{arabic}
\pagestyle{myheadings}
\thispagestyle{empty}
{\flushleft\includegraphics[width=\textwidth,bb=58 650 590 680]{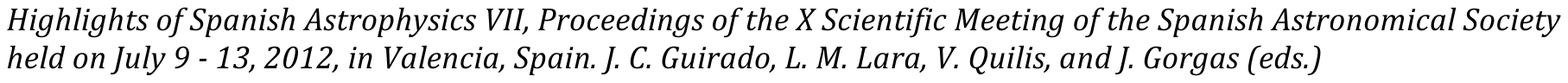}}
\vspace*{0.2cm}
\begin{flushleft}
{\bf {\LARGE
%
%%% TITLE of the paper. 
%%% TITLE of the paper. 
Using radiative transfer models to study the atmospheric water vapor content and to eliminate telluric lines from high-resolution optical spectra
%
% Do not delete next few lines
}\\
\vspace*{1cm}
%
%%% Include here the LIST OF AUTHORS.
%%% Include here the LIST OF AUTHORS.
%%% Note that the last author has to be preceeded by an AND.
A. Gardini$^{1}$,
J. Ma\'{\i}z Apell\'aniz$^{1}$,
E. P\'erez$^{1}$,
J. A. Quesada,
and
B. Funke$^{1}$ 
%
% Do not delete next few lines
}\\
\vspace*{0.5cm}
%
%%% AFFILIATIONS LIST.
%%% and the AFFILIATIONS LIST. Note that one affiliation per line.
%%% Add as many affiliations as necessary. 
$^{1}$
Instituto de Astrof\'{\i}sica de Andaluc\'{\i}a-CSIC, Glorieta de la Astronom\'{i}a s/n, \linebreak 18008 Granada, Spain
%
% Do not delete next few lines
\end{flushleft}
%
% Headings
\markboth{
%%% Type the SHORT version of the paper title.
%%% Type the SHORT version of the paper title.
Using RTM to study the WV content and eliminate telluric lines from high-resolution optical spectra
}{ % Do not delete
%
%%%  First Author \& Second Author   OR   First-author et al. 
%%%  First Author \& Second Author   OR   First-author et al. if the author list 
%%% contains three or more authors.
Gardini et al.
% 
% Do not delete next few lines
}
\thispagestyle{empty}
\vspace*{0.4cm}
\begin{minipage}[l]{0.09\textwidth}
\ 
\end{minipage}
\begin{minipage}[r]{0.9\textwidth}
\vspace{1cm}
\section*{Abstract}{\small
%
% ABSTRACT ABSTRACT ABSTRACT
% ABSTRACT ABSTRACT ABSTRACT
%%% Type the ABSTRACT of your paper
The Radiative Transfer Model (RTM) and the retrieval algorithm, incorporated in the SCIATRAN 2.2 software package developed at the Institute of 
Remote Sensing/Institute of Enviromental Physics of Bremen University (Germany), allows to simulate, among other things, radiance/irradiance 
spectra in the 2400-24\,000 \AA\ range. In this work we present applications of RTM to two case studies. In the first case the RTM was used to 
simulate direct solar irradiance spectra, with different water vapor amounts, for the study of the water vapor content in the atmosphere above 
Sierra Nevada Observatory. Simulated spectra were compared with those measured with a spectrometer operating in the 8000-10\,000 \AA\ range. 
In the second case the RTM was used to generate telluric model spectra to subtract the atmospheric contribution and correct high-resolution 
stellar spectra from atmospheric water vapor and oxygen lines. The results of both studies are discussed.

%
% Do not delete next few lines
\normalsize}
\end{minipage}
%
%
%%% BODY of the paper
%%% BODY of the paper

\section{A simple idea}

\begin{itemize}
 \item Use modern radiative transfer models of the Earth's atmosphere for astronomical applications.
 \item Two uses:
 \begin{itemize}
  \item Site testing for water vapor content.
  \item Elimination of telluric lines from optical spectra without the need for contemporary telluric standards.
 \end{itemize}
\end{itemize}

\begin{figure}
\centerline{
\includegraphics[width=0.92\linewidth]{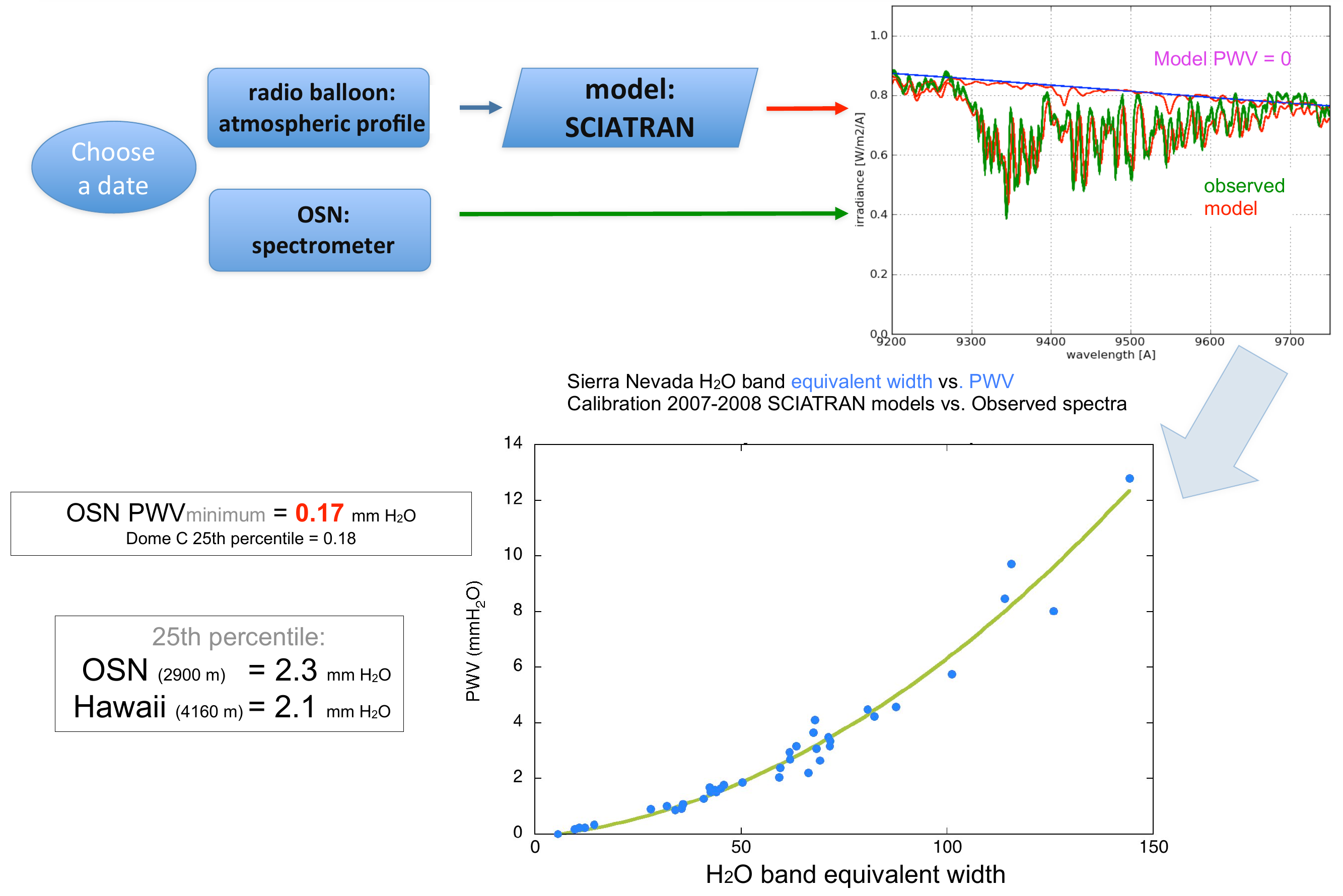}
}
\caption{The extremely dry atmosphere above Sierra Nevada: an excellent calibration.}
\end{figure}

\section{The software}

$\,\!$\indent The Radiative Transfer Model (RTM) and the retrieval algorithm, incorporated in the SCIATRAN 2.2 software package developed at the Institute of Remote
Sensing/Institute of Enviromental Physics of Bremen University (Germany), allows to simulate, among other things, radiance/irradiance spectra in the 
2400~-~24\,000~\AA\ range. In this work we present the applications of RTM to two case studies. In the first case the RTM was used to simulate direct solar irradiance
spectra, with different water vapor amounts, for the study of the water vapor content in the Sierra Nevada Observatory (OSN) atmosphere. Simulated
spectra were compared with those measured with a spectrometer operating in the 8000~-~10\,000~\AA\ range.
In the second case the RTM was used to generate a telluric model to subtract the atmospheric contribution and correct high-resolution stellar spectra from atmospheric 
water vapor and molecular oxygen lines. The results of both cases are discussed here.

\begin{figure}
\centerline{
\includegraphics[width=0.79\linewidth]{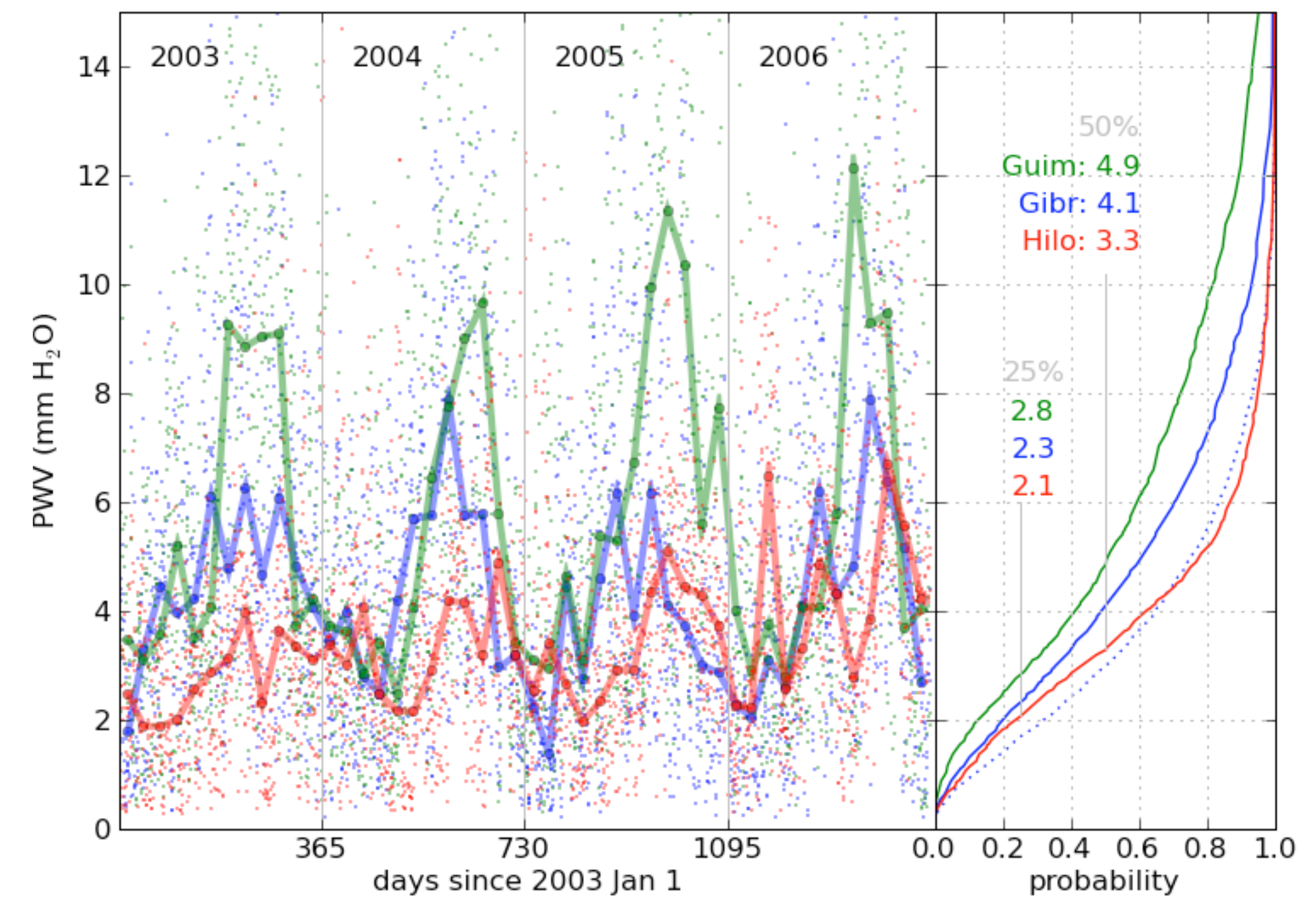}
}
\caption{2003 - 2006 data (left) and statistics (right) of the radio balloon measurements from Canarias (proxy G\"uimar, green), Granada (proxy Gibraltar, blue), 
and Hawaii (proxy Hilo, red). In the left panel dots are individual measurements and lines are averages.}
\end{figure}

\section{A study of the water vapor content at the Sierra Nevada Observatory}

$\,\!$\indent The OSN is the southernmost high altitude (2900 m a.s.l.) location in continental Europe. It is located in Sierra Nevada, Granada, Spain 
(37.06 N, 3.38 W), less than one km away from the IRAM 30 m telescope. Given the dry climatic conditions it can be considered a very competitive location for 
MIR-submm astronomical observations.  We obtained direct solar irradiance spectra with an array spectrometer in the 8000~-~10\,000~\AA\ range (2.5~\AA\ spectral 
resolution), during the 2007 - 2009 campaign, for a total of 248 days. The Wyoming University bi-daily radiosounding data were used to provide profiles of 
Temperature (T), Pressure (P) and H$_2$O volumetric mixing ratio (vmr) as a function of the geopotential height. The 2003 - 2006 statistics of the radio balloon 
measurements from Granada (proxy Gibraltar), Canarias (proxy G\"uimar), and Hawaii (proxy Hilo) shows that OSN is a very competitive site (Figs.~1~and~2). The RTM 
was used to simulate the direct solar irradiance spectra at OSN and to compare them with those measured with the spectrometer. The equivalent width (EW) of measured 
spectra in the same water vapor absorption band were also computed.

\begin{figure}
\centerline{
\includegraphics[width=0.92\linewidth, bb=28 28 566 351]{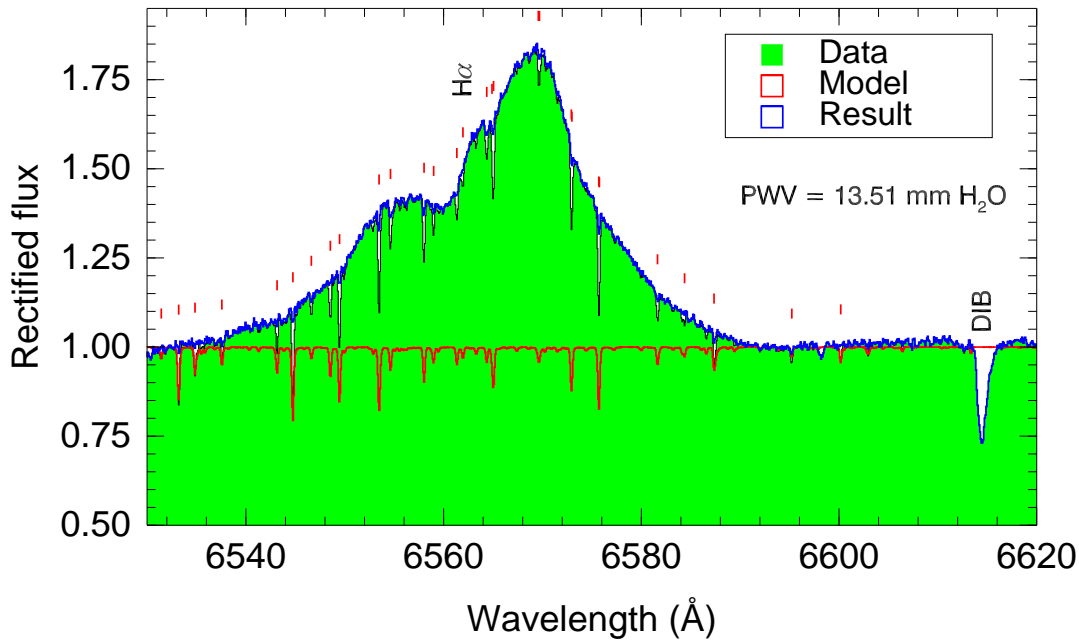}
}
\caption{Elimination of H$_2$O telluric lines in the H$\alpha$ region. Data obtained with HRS/HET at $R = 30\,000$. 
The red vertical bars mark the location of the strongest telluric lines in this wavelength range.}
\end{figure}

We used only the spectra observed $\pm$1 hour from the Gibraltar noon radiosounding (12:00 Z). For each day, the profiles of P, T and H$_2$O vmr provided
by the radiosounding were used to define the model atmosphere in the RTM. For each simulated spectrum the EW in the 9270~-~9700~\AA\ water vapor absorption
band was computed, and the corresponding PWV above OSN was associated to it. This provides the calibration shown in Fig. 1 (bottom), where blue dots are the 
individual models for 2007 - 2009 campaign, and the green line is a second order polynomial fit that goes through the ``origin'' EW. The ``origin'' EW is 
measured for an atmosphere of zero H$_2$O column and contributed by the other residual gases (see the low EW pink line in the spectrum above). Notice that we have 
measured very dry epochs with PWV $<$ 0.2 mm H$_2$O. We are carrying out an observing campaign to quantify in more detail the statistics of very dry epochs, 
PWV $<$ 1 mm H$_2$O, that are certainly not uncommon in Sierra Nevada.

\begin{figure}
\centerline{
\includegraphics[width=0.82\linewidth, bb=28 28 566 351]{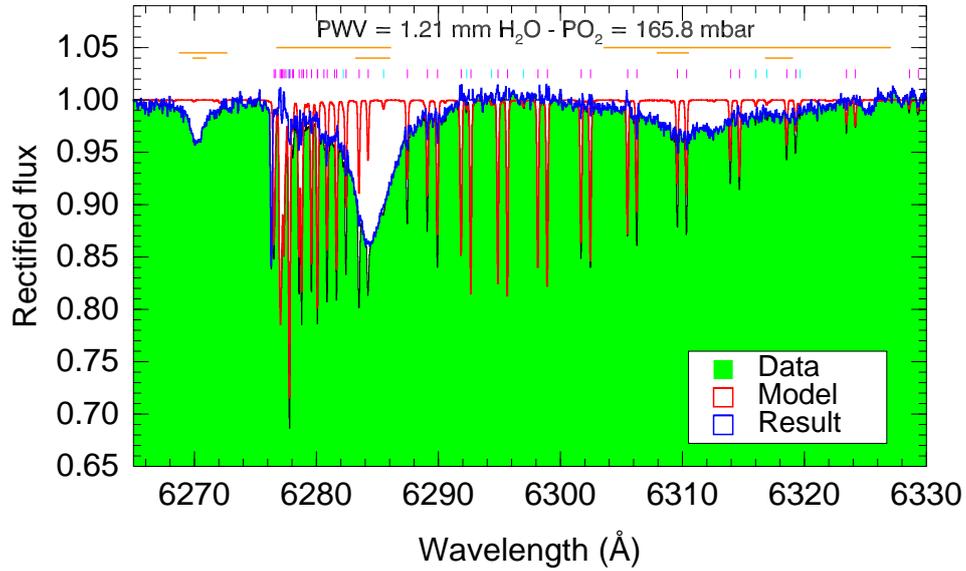}
}
\caption{Elimination of H$_2$O and O$_2$ telluric lines in the 6300 \AA\ region. Data obtained with FEROS at $R = 48\,000$. The vertical bars mark the location of 
the weak H$_2$O (cyan) and strong O$_2$ (magenta) telluric lines. The horizontal orange bars indicate the location and approximate widths of 
the diffuse interstellar bands (DIBs) in this region. Note that the model does not provide an accurate fit for the shortest wavelength (6277 \AA) telluric line.}
\end{figure}

\section{Elimination of telluric lines from high-resolution optical spectra}

\begin{itemize}
 \item We generate two reference models for 3050 - 10\,000 \AA.
 \begin{itemize}
  \item Obtained for the Sierra Nevada Observatory ($h$ = 2900 m, $P_{\rm O_2}$ = 151 mbar) but applicable anywhere after rescaling.
  \item Only contribution from lines in the output at very high ($R > 100\,000$) spectral resolution.
  \item One model without H$_2$O (``oxygen model'').
  \item One model with 2.78 mm of H$_2$O (``oxygen + water model'').
  \end{itemize}
 \item IDL code to fit the telluric spectrum without contemporary telluric standards (Figs.~3~to~6).
 \begin{itemize}
  \item Two (O$_2$ + H$_2$O) times three (column, velocity, spectral resolution) free parameters.
  \item Possibility of defining different wavelength ranges to account for resolution differences in the spectrum.
  \item Tested with $R = 30\,000 - 85\,000$ spectra from Hermes (Mercator, La Palma), FIES (NOT, La Palma), FEROS (2.2 m, La Silla), and HRS (HET, McDonald).
 \end{itemize}
\end{itemize}

\begin{figure}
\centerline{
\includegraphics[width=1.20\linewidth, bb=28 28 566 458]{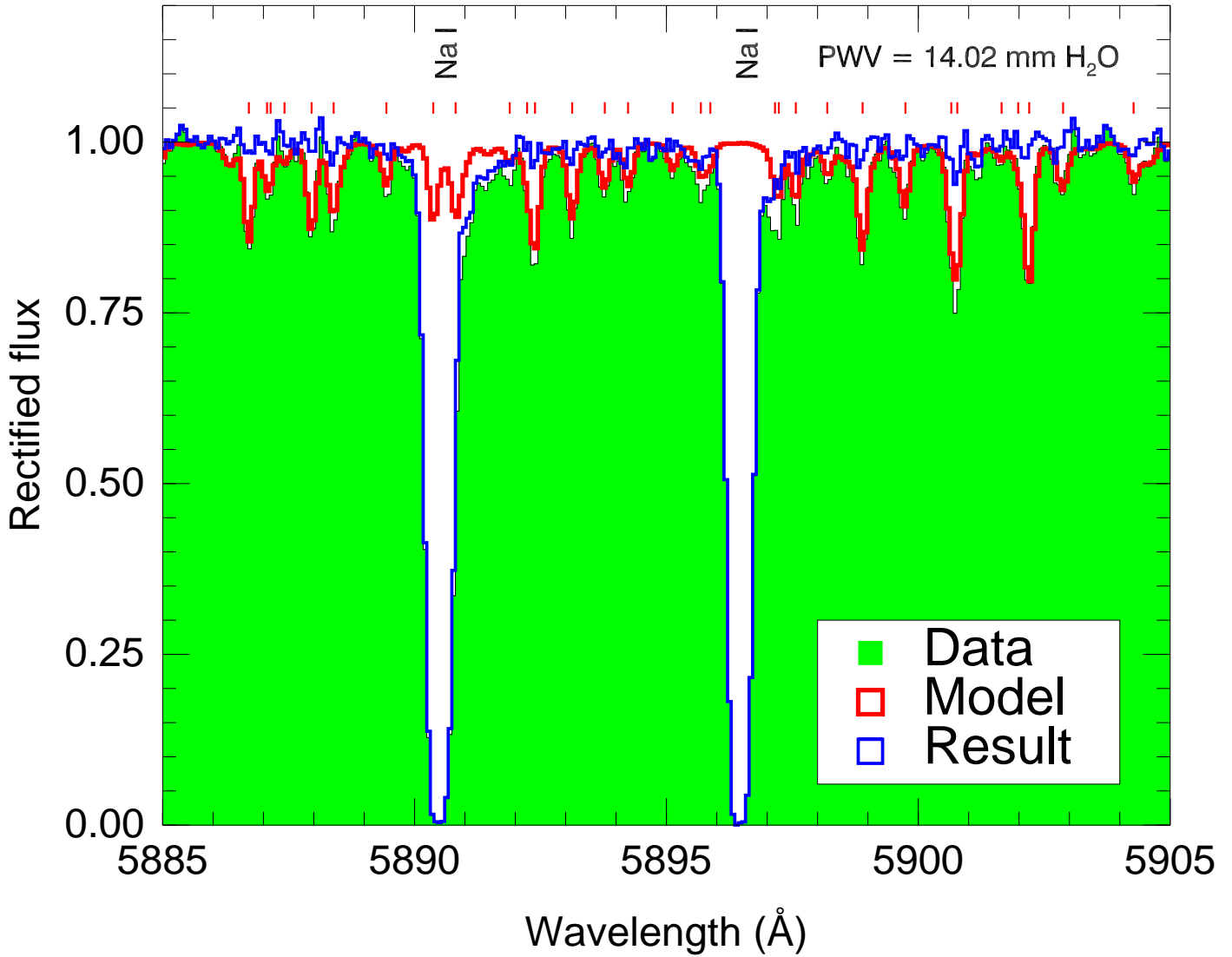}
}
\caption{Elimination of H$_2$O telluric lines in the region of the Na\,{\sc i}~D1+D2 interstellar doublet. Data obtained with HRS/HET at $R = 30\,000$. The red 
vertical bars mark the location of the strongest telluric lines in this wavelength range. Note the humid conditions under which the spectrum was obtained.}
\end{figure}

\begin{figure}
\centerline{
\includegraphics[width=1.20\linewidth, bb=28 28 566 458]{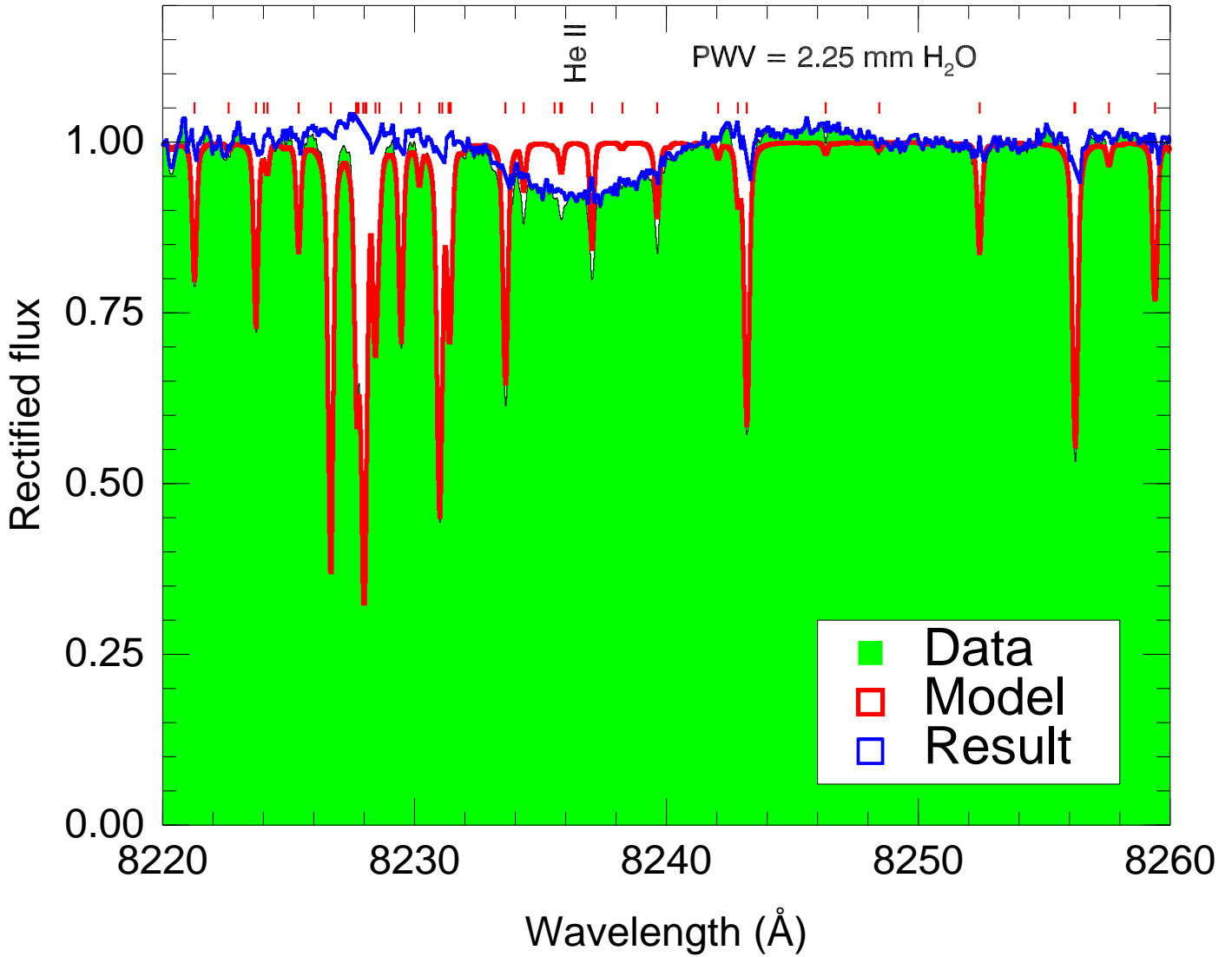}
}
\caption{Elimination of H$_2$O telluric lines in the region of the He\,{\sc ii}~$\lambda$8237~\AA\ stellar absorption line. 
This He\,{\sc ii} line is the best identifier for O stars in the 5500~-~10\,000~\AA\ range and is located in the middle of a strong H$_2$O telluric band. 
Data obtained with FEROS at $R = 48\,000$. The red vertical bars mark the location of the strongest telluric lines in this wavelength range.}
\end{figure}

%
%
% Do not delete the next line
\small  % Do not delete
%
%%% Comment the following line if you do not have acknowledgments.
%\section*{Acknowledgments}   % Do not delete if you declare acknowledgments
%
%%% ACKNOWLEDGMENTS
%%% ACKNOWLEDGMENTS

%
% Do not delete the next few lines
%\begin{thebibliography}{}
%\small
%
%%% BIBLIOGRAPHY
%%% BIBLIOGRAPHY
%
%
% Do not delete next few lines
%\end{thebibliography}
%
\end{document}